\documentclass[english,aps,manuscript]{revtex4-1}
\usepackage{charter}
\usepackage[T1]{fontenc}
\usepackage[latin9]{inputenc}
\usepackage[letterpaper]{geometry}
\usepackage{amsmath}
\usepackage{graphicx}
\usepackage{esint}
\usepackage{subscript}

\makeatletter
 
 \@ifundefined{textcolor}{}
 {%
   \definecolor{BLACK}{gray}{0}
   \definecolor{WHITE}{gray}{1}
   \definecolor{RED}{rgb}{1,0,0}
   \definecolor{GREEN}{rgb}{0,1,0}
   \definecolor{BLUE}{rgb}{0,0,1}
   \definecolor{CYAN}{cmyk}{1,0,0,0}
   \definecolor{MAGENTA}{cmyk}{0,1,0,0}
   \definecolor{YELLOW}{cmyk}{0,0,1,0}
 }

\makeatother

\usepackage{babel}
\begin{document}

\title{Phase separation in ion-irradiated compound semiconductors: \\
an alternate route to ordered nano-structures}

\author{Scott A. Norris}

\email{snorris@smu.edu}

\selectlanguage{english}%

\affiliation{Southern Methodist University\\
Department of Mathematics\\
}

\begin{abstract}
In recent years, observations of highly-ordered, hexagonal arrays
of self-organized nanostructures on binary or impurity-laced targets
under normal-incidence ion irradiation have excited interest in this
phenomenon as a potential route to high-throughput, low-cost manufacture
of nanoscale devices or nanostructured coatings. The currently-prominent
explanation for these structures is a morphological instability driven
by ion erosion discovered by Bradley and Shipman; however, recent
parameter estimates via molecular dynamics simulations suggest that
this erosive instability may not be active for the representative
GaSb system in which hexagonal structures were first observed.

Motivated by experimental and numerical evidence suggesting the possible
importance of phase separation in ion-irradiated compounds, we here
generalize the Bradley-Shipman theory to include the effect of ion-assisted
phase separation. The resulting system admits a chemically-driven
finite-wavelength instability that can explain the order of observed
patterns even when the erosive Bradley-Shipman instability, and in
a relevant simplifying limit, provides an intuitive instability criteria
that agrees qualitatively with experimental observations on pattern
wavelengths. Finally, we identify a characteristic experimental signature
that distinguishes the chemical and morphological instabilities, and
highlights the need for specific additional experimental data on the
GaSb system.
\end{abstract}
\maketitle

\section{Introduction}

Among the many nanoscale patterns that have been observed on ion-irradiated
surfaces \cite{chan-chason-JAP-2007,frost-etal-APA-2008}, the discovery
of hexagonal arrays of high-aspect ratio dots on Ar\textsuperscript{+}-irradiated
GaSb \cite{facsko-etal-SCIENCE-1999} has sparked a flurry of experimental
and theoretical study into spontaneous pattern formation as a potential
route to ``bottom-up'' fabrication of nanoscale devices. The mathematical
study of such patterns is a mature field known as \emph{bifurcation
theory}, and the ingredients necessary to produce long-range hexagonal
order are well known \cite{cross-hohenberg-RoMP-1993,cross-greenside-2009-book}.
First, one needs isotropic physics, which is readily achieved by irradiating
the sample at normal incidence, or by rotating it during off-normal
bombardment. Second, one needs a destabilizing mechanism that produces
a \emph{finite wavelength bifurcation} -- a transition in the linearized
equations from stability to instability that occurs first for perturbations
of a finite wavelength. This kind of bifurcation leads to instabilities
in which the unstable wavenumbers are bounded both above and below
by stable wavenumbers; the resulting narrow band of unstable modes
generically produces some kind of ordered pattern at long times.

Unfortunately, none of the early models of ion irradiation predicted
a narrow-band instability \cite{davidovitch-etal-PRB-2007,madi-etal-2008-PRL},
suggesting a need for the incorporation of additional physical effects.
Two candidates that initially seemed promising were a ``damping''
term proposed to stand for long-range atomic redeposition \cite{facsko-etal-PRB-2004}
and biaxial compressive stress injected by the ion beam (see \cite{davidovitch-etal-PRB-2007,madi-etal-2008-PRL}
for a comprehensive discussion). However, recently both mechanisms
have been ruled out as sources of a narrow band \cite{bradley-PRB-2011a,norris-PRB-2012-viscoelastic-normal}.
In parallel with these theoretical struggles, persistent difficulty
obtaining consistent results between labs has led to a re-examination
of the experimental results. It was found that dots, originally also
observed on irradiated pure silicon \cite{gago-etal-APL-2006,ziberi-etal-APL-2008},
disappeared from that system when impurities and geometric artifacts
were carefully removed \cite{madi-etal-2008-PRL,madi-etal-JPCM-2009,madi-aziz-ASS-2012},
and conversely, that the systematic addition of impurities to a clean
system causes dots to appear \cite{ozaydin-etal-APL-2005,ozaydin-etal-JVSTB-2008,ozaydin-ludwig-JPCM-2009,zhang-etal-NJoP-2011}.
This has led to the strengthening conclusion that two-component materials
are the necessary ingredient to produce ordered quantum dots \cite{macko-etal-NanoTech-2010}. 

This growing consensus has caused increasing interest in theoretical
treatments of irradiated materials with two components, in which coupled
PDEs simultaneously describe both the \emph{morphology} and the \emph{concentration}
evolution. Extending the pioneering work Shenoy, Chan, and Chason
\cite{shenoy-chan-chason-PRL-2007}, Bradley and Shipman (BS) have
recently introduced such a theory for irradiated binary compounds
exhibiting the first physically-grounded finite-wavelength bifurcation
\cite{bradley-shipman-PRL-2010,shipman-bradley-PRB-2011}, making
it a foundation for current understanding of these structures. However,
in order to explain the structures, the model makes several strong
assumptions. First, it requires that any finite-wavelength bifurcation
be driven primarily by the erosive instability identified by Bradley
and Harper \cite{bradley-harper-JVST-1988}, which implies that surfaces
should be unstable even in the absence of a second species. Second,
it requires a strong component of preferential redistribution, in
which one target atom is redistributed much more strongly than the
other. Last, it assumes chemical diffusion of the standard, Fickian
type.

As will be shown below, all three assumptions of the Bradley-Shipman
theory may be fairly questioned, with some evidence to the contrary
existing in each case. This suggests the consideration of alternate
physical mechanisms to explain the ordered structures in irradiated
binary compounds. Here, based on a growing body of experimental \cite{fukutani-etal-JJAP-2008,le-roy-etal-JAP-2009,abrasonis-etal-JAP-2010,hoffsass-etal-APA-2012}
and numerical \cite{fukutani-etal-JJAP-2008,le-roy-etal-PRB-2010-phase-separation}
evidence in a variety of irradiation regimes, we propose a generalization
of the original BS theory that admits the process of phase separation.
We find that our model admits a finite-wavelength bifurcation even
when the assumptions of the BS theory are not fulfilled; in addition,
it provides a clear experimental signature distinguishing it from
the Bradley-Shipman mechanism. In a plausible and particularly useful
limit of nearly equal redistribution of the two species, we obtain
an intuitive approximation of the instability criteria, which shows
that the presence of the ion beam fundamentally changes the nature
of the instability relative to other phase-separating systems.

\section{Brief Review of the Bradley-Shipman Model\label{sec: review-of-BS}}

In order to provide proper historical and mathematical context, and
because our treatment of phase separation builds upon it, we begin
by very briefly summarizing the recent work of Bradley and Shipman
\cite{bradley-shipman-PRL-2010,shipman-bradley-PRB-2011} for binary
compounds irradiated at normal incidence, highlighting those results
that we will refer to directly in our generalization.

\subsection{Model}

The Bradley-Shipman model tracks the evolution of a height field $z=h\left(x,y,t\right)$
describing the irradiated surface, and concentration fields $c_{A}\left(x,y,t\right)$
and $c_{B}\left(x,y,t\right)$ of two components $A$ and $B$. Under
the effects of normal-incidence ion bombardment, one material is preferentially
eroded until a steady state is reached, in which the surface is receding
with constant velocity $v_{0}$ and with constant surface concentrations
$c_{A,0}$ and $c_{B,0}$ of $A$ and $B$ atoms, respectively. Perturbations
to this steady state are then investigated by setting 
\[
\begin{aligned}h\left(x,y,t\right) & =-v_{0}t+u\left(x,y,t\right)\\
c_{A}\left(x,y,t\right) & =c_{A,0}+\phi\left(x,y,t\right)\\
c_{B}\left(x,y,t\right) & =c_{B,0}+\left(1-\phi\left(x,y,t\right)\right)
\end{aligned}
\]
where $u\left(x,y,t\right)$ and $\phi\left(x,y,t\right)$ describe
the perturbations to the height and concentration field of species
$A$. After some significant analysis (Eqs.~(3)-(13) of Ref.\cite{shipman-bradley-PRB-2011}),
and the neglect of a few terms based on physical considerations, the
authors obtain the following linearized equations for the evolution
of $u$ and $\phi$:
\begin{eqnarray}
\frac{\partial u}{\partial t} & = & -A\phi\phantom{+B\nabla^{2}\phi}+C\nabla^{2}u-D\nabla^{4}u\label{eqn: linear-height-evolution-BH}\\
\frac{\partial\phi}{\partial t} & = & -A^{\prime}\phi+B^{\prime}\nabla^{2}\phi+C^{\prime}\nabla^{2}u\phantom{-D^{\prime}\nabla^{4}u}.\label{eqn: linear-concentration-evolution-BH}
\end{eqnarray}
The coefficients in Eqn.(\ref{eqn: linear-height-evolution-BH}) for
the height evolution are defined via

\begin{equation}
\begin{aligned}A & =P_{0}\Omega\left[\Lambda_{A}^{\prime}\left(c_{A,0}\right)-\Lambda_{B}^{\prime}\left(c_{B,0}\right)\right]\\
C & =\Omega\left[\left(\mu_{A}\left(c_{A,0}\right)+\mu_{B}\left(c_{B,0}\right)\right)-\alpha\left(\Lambda_{A}\left(c_{A,0}\right)+\Lambda_{B}\left(c_{B,0}\right)\right)\right]\\
D & =\left[c_{A,0}D_{A}+c_{B,0}D_{B}\right]\frac{n_{s}\Omega^{2}\gamma_{s}}{k_{B}T}>0
\end{aligned}
\label{eq: ACD}
\end{equation}
and those in Eqn.(\ref{eqn: linear-concentration-evolution-BH}) for
the concentration evolution, via

\begin{equation}
\begin{aligned}A^{\prime} & =\frac{P_{0}\Omega}{\Delta}\left[c_{B,b}\Lambda_{A}^{\prime}\left(c_{A,0}\right)+c_{A,b}\Lambda_{B}^{\prime}\left(c_{B,0}\right)\right]>0\\
B^{\prime} & =\frac{n_{s}\Omega}{\Delta}\left[c_{B,b}D_{A}+c_{A,b}D_{B}\right]>0\\
C^{\prime} & =\frac{\Omega}{\Delta}\left[c_{B,b}\mu_{A}\left(c_{A,0}\right)-c_{A,b}\mu_{B}\left(c_{B,0}\right)\right]
\end{aligned}
.\label{eq: alpha-beta-gamma}
\end{equation}
Here $P_{0}$ is the power deposited by the ions per unit surface
area on a flat surface, $\Omega$ is the atomic volume (taken to be
the same for both species), $n_{s}$ is the total number of mobile
surface atoms per unit area on the surface, $\gamma_{s}$ is the surface
energy, $k_{B}$ is Boltzmann's constant, $T$ is the temperature,
$\Delta$ is the amorphous film thickness, $c_{A,0}$ and $c_{B,0}$
are the concentration of $A$ and $B$ atoms in the film at steady
state, $c_{A,b}$ and $c_{B,b}$ are the concentration of $A$ and
$B$ atoms in the bulk, $D_{A}$ and $D_{B}$ are the diffusivities
of $A$ and $B$ atoms, $\Lambda_{A}\left(c_{A}\right)$ and $\Lambda_{B}\left(c_{B}\right)$
are proportionality constants linking the deposited power to the sputtering
rate, and $\mu_{A}\left(c_{A}\right)$ and $\mu_{B}\left(c_{B}\right)$
are proportionality constants describing preferential redistribution
of material, an effect first proposed in the Bradley-Shipman model,
and playing a critical role therein.

We have continued to employ the assumption of BS that the mobilities
of the two species are equal, which eliminates a term $B\nabla^{2}c$
in Eqn.(\ref{eqn: linear-height-evolution-BH}). In addition, those
authors neglect a term $D^{\prime}\nabla^{4}h$ in Eqn.(\ref{eqn: linear-concentration-evolution-BH}),
because it does not seem to affect numerical simulations much. (If
one replaces surface diffusion with ion-enhanced viscous flow \cite{umbach-etal-PRL-2001}
as the dominant surface energy relaxation mechanism, this neglect
becomes rigorous.) In a slight deviation from the definitions in the
BS model, we have defined coefficients in such a way that zeroth-
and fourth-order terms in Eqs.(\ref{eqn: linear-height-evolution-BH})-(\ref{eqn: linear-concentration-evolution-BH})
have negative sign, whereas second-order terms have positive sign.
In an additional slight deviation, we assume without loss of generality
that the species labels A and B are chosen in such a way that $\Lambda_{A}^{\prime}\left(c_{A,0}\right)>\Lambda_{B}^{\prime}\left(c_{B,0}\right)$,
so that the coefficient $A$ in Eq.\ref{eqn: linear-height-evolution-BH}
is positive. This will greatly simplify a discussion of experimental
signatures in Section~\ref{sec: experimental-signature} below.

\subsection{Stability Analysis\label{sub: Stability-Analysis}}

Because the surface is irradiated at normal incidence, the system
is isotropic, with no preferred orientation. Therefore, to determine
the presence or absence of an instability, and its general characteristics,
it is sufficient to consider a one-dimensional perturbation. (The
ultimate long-time morphology of the pattern, of course, depends on
careful study of superimposed modes \cite{cross-greenside-2009-book,bradley-shipman-PRL-2010,shipman-bradley-PRB-2011}
in the presence of nonlinearities.) Without loss of generality, one
may orient this instability in the $x$-direction, giving

\begin{equation}
\left[\begin{array}{c}
u_{\phantom{1}}\\
\phi_{\phantom{1}}
\end{array}\right]=\left[\begin{array}{c}
u_{1}\\
\phi_{1}
\end{array}\right]e^{ikx+\sigma\left(k\right)t},\label{eqn: infinitesimal-perturbation}
\end{equation}
where $k$ is the wavenumber of the perturbation, $\sigma\left(k\right)$
the growth rate of that perturbation, and $h_{1}$ and $\phi_{1}$
are undetermined constants describing the relative phases and magnitudes
of the height and concentration modulations. Inserting the perturbation
(\ref{eqn: infinitesimal-perturbation}) into the governing equations
(\ref{eqn: linear-height-evolution-BH})-(\ref{eqn: linear-concentration-evolution-BH}),
one obtains, in matrix form,
\begin{equation}
\left[\begin{array}{cc}
\sigma+Ck^{2}+Dk^{4} & A\\
C^{\prime}k^{2} & \sigma+A^{\prime}+B^{\prime}k^{2}
\end{array}\right]\left[\begin{array}{c}
u_{1}\\
\phi_{1}
\end{array}\right]=\mathbf{0}.\label{eqn: stability-matrix-BS}
\end{equation}
 For a solution $\left[u_{1},\,\phi_{1}\right]^{T}$ of this equation,
the determinant of the matrix must be zero, giving a the dispersion
relation $\sigma\left(k\right)$ in the quadratic form 
\begin{equation}
\sigma^{2}+\tau\sigma+\Delta=0,\label{eqn: quadratic-equation-BS}
\end{equation}
where $\tau\left(k\right)$ and $\Delta\left(k\right)$ are respectively
the trace and determinant of (\ref{eqn: stability-matrix-BS}) when
$\sigma=0$:

\begin{eqnarray}
\tau\left(k\right) & = & A^{\prime}+\left(C+B^{\prime}\right)k^{2}+Dk^{4}\label{eqn: tau-BS}\\
\Delta\left(k\right) & = & \left(CA^{\prime}-AC^{\prime}\right)k^{2}+\left(CB^{\prime}+DA^{\prime}\right)k^{4}+\left(DB^{\prime}\right)k^{6}.\label{eqn: Delta-BS}
\end{eqnarray}

For the purpose of identifying stability properties, it is sufficient
to consider only the root $\sigma_{+}\left(k\right)$ associated with
the '+' sign in the quadratic formula, which gives the growth rate
of the faster-growing solution:
\begin{equation}
\sigma_{+}\left(k\right)=\frac{1}{2}\left(-\tau+\sqrt{\tau^{2}-4\Delta}\right).\label{eqn: quadratic-solution-of-sigma-BS}
\end{equation}
The value of $k$ that maximizes $\sigma\left(k\right)$ - denoted
$k_{\text{max}}$ - is called the \emph{most unstable mode}; if $\sigma\left(k_{\text{max}}\right)<0$,
then all modes decay and the steady solution is stable, whereas if
$\sigma\left(k_{\text{max}}\right)>0$, then at least some modes grow
and the steady solution is unstable. A set of parameters at which
$\sigma\left(k_{\text{max}}\right)$ changes from negative to positive
is called a \emph{bifurcation point}. If the bifurcation occurs for
some finite value $k_{\text{max}}\ne0$, the bifurcation is said to
be \emph{finite-wavelength}, since the most unstable mode has a finite
wavelength of $\lambda=2\pi/k_{\text{max}}$. However, if it occurs
for $k_{\text{max}}=0$, it is said to be \emph{longwave}, because
the wavelength of the most unstable mode approaches infinity. Beyond
the bifurcation point, finite-wavelength bifurcations exhibit a narrow
band of unstable modes, with stable modes at both larger and smaller
values of $k$. This leads generically to ordered arrays of stripes,
squares, or hexagons. On the other hand, longwave bifurcations tend
to exhibit instability for all wavelengths larger than a critical
value, leading to disordered kinetic roughening phenomena \cite{cross-hohenberg-RoMP-1993,cross-greenside-2009-book}.

\subsection{Discussion: Requirements for a Finite-Wavelength Bifurcation\label{sub: fwb-requirements}}

A full analysis of Eqn.(\ref{eqn: quadratic-solution-of-sigma-BS})
is contained in Refs.\cite{bradley-shipman-PRL-2010,shipman-bradley-PRB-2011}.
Here, we obtain only a few results relevant to what follows, in a
rigorous but qualitative way emphasizing intuition. In particular,
an instability with a narrow band of unstable modes must have stable
modes as in both of the limits $k\to0$ and $k\to\infty$, so these
limits merit brief examination.

In fact, all well-posed physical problems should exhibit $\lim_{k\to\infty}\sigma_{+}\left(k\right)<0$;
i.e., that very-small scale perturbations are stable. Otherwise, the
system would be expected to exhibit variation on infinitely-small
scales, violating the continuum hypothesis and suggesting that additional
physical terms were necessary in the model. From Equations~(\ref{eqn: tau-BS})-(\ref{eqn: quadratic-solution-of-sigma-BS})
a simple asymptotic expansion reveals that 
\begin{equation}
\lim_{k\to\infty}\sigma_{+}\left(k\right)\approx-2\left(DB^{\prime}\right)k^{2}+\mathcal{O}\left(1\right).\label{eqn: large-k-limit}
\end{equation}
Because it is assumed that $D>0$ we see that equations (\ref{eqn: linear-height-evolution-BH})-(\ref{eqn: linear-concentration-evolution-BH})
are well-posed as long as $B^{\prime}>0$, which is also assumed within
the model. This amounts to an assumption that chemical diffusion is
of the standard, Fickian type.

To achieve a narrow band of unstable wavenumbers, it is also necessary
that the long waves in the limit $k\to0$ be stable. Again, a simple
asymptotic expansion of Eqns.~(\ref{eqn: tau-BS})-(\ref{eqn: quadratic-solution-of-sigma-BS})
in the limit as $k\to0$ reveals that
\begin{equation}
\lim_{k\to0}\sigma_{+}\left(k\right)\approx-\frac{1}{A^{\prime}}\left(CA^{\prime}-AC^{\prime}\right)k^{2}.\label{eqn: small-k-limit-1}
\end{equation}
This leads immediately to a requirement for narrow-band instabilities
of

\begin{equation}
CA^{\prime}>AC^{\prime}.\label{eqn: finite-wavelength-requirement}
\end{equation}

A general feature of Eq.(\ref{eqn: quadratic-solution-of-sigma-BS})
is that a mode is only unstable if either $\tau<0$ or $\Delta<0$
\cite{cross-greenside-2009-book}. Together with the limits just considered,
this allows us to easily summarize the main assumptions that the BS
theory requires in order to explain ordered structures. First, under
the restriction (\ref{eqn: finite-wavelength-requirement}), and because
the parameters $D$, $A^{\prime}$, and $B^{\prime}$ are all positive
by definition, we see that the only way to achieve $\tau<0$ or $\sigma<0$
is if the coefficient $C$ is sufficiently negative to drive the instability.
This is a notable criteria, because $C<0$ implies the existence of
an erosion-driven morphological instability \cite{bradley-harper-JVST-1988}
even in the absence of concentration modulations ($\phi_{1}\equiv0$).
Second, if the coefficient $C$ must be negative, then to satisfy
Eq.(\ref{eqn: finite-wavelength-requirement}), the parameter group
$AC^{\prime}$ must be sufficiently negative to stabilize the long
waves, with $\left|AC^{\prime}\right|>\left|A^{\prime}C\right|$.
This implies a critical role to be played by preferential redistribution,
as characterized by the parameter $C^{\prime}$, in which one atomic
species exhibits much larger displacements within the collision cascade
\cite{bradley-shipman-PRL-2010,shipman-bradley-PRB-2011}.

\section{Generalization to include phase separation\label{sub: Generalization}}

The preceding summarized analysis clearly highlights the essential
requirements of the Bradley-Shipman theory for producing ordered patterns:
(a) an erosion-driven instability indicated by $C<0$, (b) strong
preferential redistribution indicated by $\left|AC^{\prime}\right|>\left|A^{\prime}C\right|$,
and (c) ordinary Fickian chemical diffusion indicated by $B^{\prime}>0$.
We now illustrate why each of these assumptions might be questioned,
and in response, propose a generalization of the BS theory which admits
the process of phase separation.

\subsection{Motivation}

Elsewhere \cite{norris-etal-arXiv-2013-coefficient-measurements},
we have used molecular dynamics simulations to estimate the values
of $\left\{ A,C,A^{\prime},C^{\prime}\right\} $ for the representative
and much-studied GaSb system, by extending the theory of crater functions
\cite{norris-etal-2009-JPCM,norris-etal-NCOMM-2011} to the regime
of binary materials. Our simulations suggest that, contrary to the
requirements of the BS theory just described, the coefficient $C$
is positive, implying the absence of the erosive morphological instability
of Bradley and Harper. In addition, we found that the parameter group
$\left|AC^{\prime}\right|\ll\left|A^{\prime}C\right|$, suggesting
that preferential redistribution plays a negligible role in the dynamics
of this system. Indeed, for the parameters we estimated, all terms
in Equations~(\ref{eqn: tau-BS})-(\ref{eqn: Delta-BS}) are positive,
which means the theory should in fact predict stable, flat surfaces
instead of the ordered structures that are frequently observed. This
provides strong motivation for the consideration of alternative instability
mechanisms to drive these patterns.

\begin{figure}
\centering{}\includegraphics[width=2.5in]{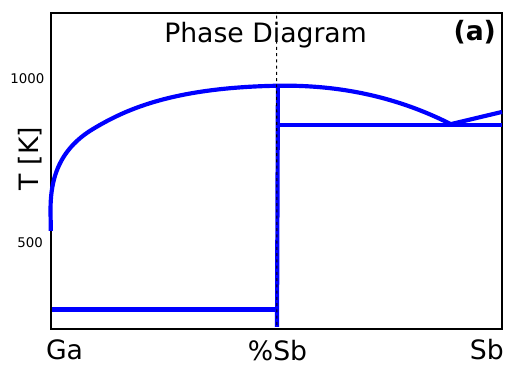}\includegraphics[width=2.5in]{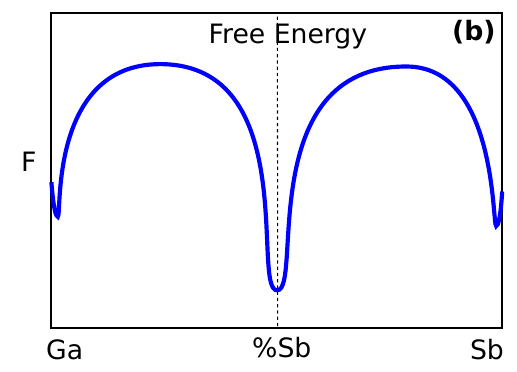}\caption{(a) The phase diagram of crystalline GaSb, from \cite{hansen-elliot-shunk-1958-binary-alloys}.
(b) Schematic of a plausible free energy function for amorphous GaSb
at low temperatures, based on (a).\label{fig: phase-diagram-and-free-energy}}
\end{figure}

Our starting point is to note that Ref.\cite{bradley-shipman-PRL-2010,shipman-bradley-PRB-2011},
and all subsequent models to date, assume the solute flux due to diffusion
is taken to be the simple Fickian one. However, there is ample evidence
to suggest that this is not always the case. For GaSb specifically,
the presence of the line compound at the 50/50 composition line in
the phase diagram depicted in Fig~\ref{fig: phase-diagram-and-free-energy}a
suggests that a free-energy digram of amorphous GaSb at low temperatures
would have a qualitative shape like the curve depicted in Fig~\ref{fig: phase-diagram-and-free-energy}b.
For such a free energy, any significant preferential sputtering of
one component will rapidly leave the film in an energetically unfavorable
configuration, in which chemical energy would be reduced by phase
separating into regions of 50/50 GaSb, and smaller regions of the
excess component in nearly pure form. Normally, the low temperatures
typically associated with ion irradiation are far below those needed
for thermal segregation, but it is known that the ion beam induces
an effective mobility in the target atoms that is many orders of magnitude
higher than for an un-irradiated target \cite{umbach-etal-PRL-2001}.
Indeed, numerical phase-field simulations for irradiated GaSb exhibited
structures qualitatively similar to those seen experimentally \cite{le-roy-etal-PRB-2010-phase-separation}
(a pattern repeated in other irradiation regimes \cite{fukutani-etal-JJAP-2008}),
but so far phase separation during ion bombardment has not been explored
analytically. This strongly motivates the development and exploration
of a model incorporating phase separation.

\subsection{Model}

Though perhaps unable to explain pattern formation in GaSb due to
parameter values outside of the pattern-forming regime, the Bradley-Shipman
model powerfully demonstrates the ability of the coupling between
height and composition displayed in Eqs.(\ref{eqn: linear-height-evolution-BH})-(\ref{eqn: linear-concentration-evolution-BH})
to admit a variety of instability types. The increased complexity
of the dispersion relation finally provided a way for the emergence
of a finite-wavelength bifurcation, overcoming a fundamental barrier
in all models of pure materials \cite{madi-etal-2008-PRL}. Therefore,
to study phase separation in irradiated systems, we will not define
an entirely new model, but will instead generalize the existing model
of Bradley and Shipman \cite{bradley-shipman-PRL-2010,shipman-bradley-PRB-2011}
using well-known results from the field of phase separation \cite{cahn-hilliard-1958:ch-equation}. 

We begin by assuming that the ion beam creates a single amorphous
phase with a free energy in the standard form
\begin{equation}
E\left(c\right)=\int_{S}\left[F\left(c\right)+\frac{\tilde{\varepsilon}}{2}\left|\nabla c\right|^{2}\right]dS,\label{eqn: chemical-energy}
\end{equation}
where $F\left(c\right)$ is a concentration-dependent Gibbs free energy
of mixing qualitatively like that in Figure~(\ref{fig: phase-diagram-and-free-energy})b,
and $\tilde{\varepsilon}$ describes a phase boundary energy that
penalizes sharp concentration gradients. From here, a local chemical
potential $\mu\left(c\right)$ is defined variationally as 
\begin{equation}
\mu\left(c\right)=\frac{\delta E}{\delta c}=F^{\prime}\left(c\right)-\tilde{\varepsilon}\nabla^{2}c.\label{eqn: chemical-potential}
\end{equation}
Next, we assume a diffusive flux $\mathbf{j}_{D}$ of species $c$
in the direction of steepest decrease in the chemical potential;
\begin{equation}
\mathbf{j}_{D}=-\tilde{D}\nabla\mu,\label{eqn: diffusive-flux}
\end{equation}
where $\tilde{D}$ is an ion-enhanced diffusional mobility. Finally,
conservation of mass leads to terms in the concentration evolution
equation of the form
\begin{equation}
\frac{\partial c}{\partial t}\propto-\tilde{D}\nabla\mathbf{j}_{D}=\tilde{D}\nabla^{2}\left[F^{\prime}\left(c\right)-\tilde{\varepsilon}\nabla^{2}c\right].\label{eqn: concentration-terms}
\end{equation}

With this generalization of the diffusive flux, the (linearized) equations
governing the evolution of perturbations $u$ and $\phi$ to height
and concentration fields become

\begin{eqnarray}
\frac{\partial u}{\partial t} & = & -A\phi\phantom{+B\nabla^{2}\phi}+C\nabla^{2}u-D\nabla^{4}u\phantom{-E\nabla^{4}\phi}\label{eqn: linear-height-evolution-PS}\\
\frac{\partial\phi}{\partial t} & = & -A^{\prime}\phi+B^{\prime}\nabla^{2}\phi+C^{\prime}\nabla^{2}u\phantom{-D^{\prime}\nabla^{4}u}-E^{\prime}\nabla^{4}\phi,\label{eqn: linear-concentration-evolution-PS}
\end{eqnarray}
which are nearly identical to Eqs.(\ref{eqn: linear-height-evolution-BH})-(\ref{eqn: linear-concentration-evolution-BH}),
except that a term proportional to $\nabla^{4}\phi$ has been added
to Eqn.(\ref{eqn: linear-concentration-evolution-PS}). In addition,
most of the coefficients of these equations are unchanged from the
prior values in Eqs.(\ref{eqn: linear-height-evolution-BH})-(\ref{eqn: linear-concentration-evolution-BH}),
except that the coefficients $B^{\prime}$ and $E^{\prime}$ are redefined
via

\begin{eqnarray}
B^{\prime} & =\frac{\Omega}{\Delta} & \tilde{D}F^{\prime\prime}\left(c_{0}\right)\label{eq: new-beta}\\
E^{\prime} & =\frac{\Omega}{\Delta} & \tilde{D}\tilde{\varepsilon},\label{eq: new-epsilon}
\end{eqnarray}
where for simplicity we have followed Eq.(11) of Ref.\cite{shipman-bradley-PRB-2011},
which assumes that diffusion is constrained to the surface. We note
that simple Fickian diffusion, as employed in the Bradley-Shipman
model \cite{bradley-shipman-PRL-2010,shipman-bradley-PRB-2011}, is
recovered by this model if $\tilde{\varepsilon}=0$ and $F\left(c\right)=\frac{1}{2}c^{2}$,
in which case the chemical potential is simply proportional to the
concentration. However, we see from Eqn.(\ref{eq: new-beta}) that
if $F^{\prime\prime}\left(c_{0}\right)<0$, diffusion will now generate
fluxes \emph{up} the gradient of concentration - this is precisely
the regime of phase separation.

\subsection{Potential Application to other regimes}

Finally, although we have primarily in mind here the GaSb system
because of the availability of parameter estimates for that system
\cite{norris-etal-arXiv-2013-coefficient-measurements}, various modifications
of the equations derived by Bradley and Shipman to describe this system
have subsequently also been employed to describe related irradiation
regimes - for instance, those describing simultaneous co-deposition
of contaminant \cite{bradley-PRB-2011c,bradley-JAP-2012}, or the
(net growth) regime of ion beam assisted deposition (IBAD) where the
total deposition rate of all atomic species exceeds the sputter rate
due to ion irradiation \cite{abrasonis-morawetz-PRB-2012}. However,
like the original Bradley-Shipman model on which they are based, these
models all assume simple Fickian diffusion. It is therefore natural
to wonder whether these additional regimes might profitably be generalized
to the phase separation case.

In fact, there is strong experimental evidence of phase separation
in each regime (which also partly motivates the present inquiry).
For instance, in the regime of erosion with simultaneous co-deposition
of metal impurities, Hoffsass et al. \cite{hoffsass-etal-APA-2012}
have recently demonstrated a strong correlation between pattern formation,
and the affinity of the metal for silicon; metals forming a MeSi\textsubscript{2}
phase (high affinity) lead readily to patterns, metals forming a MeSi
phase (moderate affinity) less readily, and metals without a silicide
phase led to no patterns. Many of these metals have similar masses
to each other, which points to the critical role of chemistry in these
samples, as opposed to kinetics. Alternatively, in the regime of ion-assisted
deposition of binary materials, a growing body of experimental evidence
has documented the formation of 3D structures including vertically-organized
alternating material layers \cite{abrasonis-etal-JAP-2010} and laterally-organized
nanowires \cite{fukutani-etal-JJAP-2008}. These structures are consistent
with numerical studies of phase separation during film growth \cite{lu-etal-PRL-2012}
or at moving fronts more generally \cite{foard-wagner-PRE-2012}.
In addition, the regime of ion-assisted growth is of particular interest
because the stoichiometry may be chosen a priori to be within the
domain of negative values of $F^{\prime\prime}$. This removes the
need for preferential sputtering to first push the material into an
unstable ratio before phase separation can occur. 

A notable feature of the generalization in the previous section is
that it only changes the definition of the coefficient $B^{\prime}$
describing diffusion, and adds a regularizing term $-E^{\prime}\nabla^{4}\phi$
to Eq.\ref{eqn: linear-concentration-evolution-PS}. In contrast,
the constants $\left\{ A,C,D,A^{\prime},C^{\prime}\right\} $ remain
unchanged by this process. Hence, the models in references \cite{bradley-PRB-2011c,abrasonis-morawetz-PRB-2012}
could be generalized in the same way with minimal changes to the respective
derivations, resulting in equations with much the same structure as
Eqs.(\ref{eqn: linear-height-evolution-PS})-(\ref{eqn: linear-concentration-evolution-PS}).

\section{Stability Analysis}

We now present a stability analysis of Eqs.(\ref{eqn: linear-height-evolution-PS})-(\ref{eqn: linear-concentration-evolution-PS}).
After briefly repeating the qualitative analysis of Section~(\ref{sec: review-of-BS}),
we obtain a full bifurcation criteria for the present model, followed
by a simplified criteria in the limit of small preferential redistribution.
This limit gives intuitive insight into the cause of instability,
and highlights several unique features of the ion-assisted phase separation
system relative to thermal phase separation.

\subsection{Qualitative Analysis: the presence of a finite-wavelength bifurcation}

The qualitative, intuition-building analysis of Section~(\ref{sub: Stability-Analysis})
may be briefly repeated. As before, inserting the standard ansatz
(\ref{eqn: infinitesimal-perturbation}) into Eqs.(\ref{eqn: linear-height-evolution-PS})-(\ref{eqn: linear-concentration-evolution-PS})
we again obtain a quadratic equation for the growth rate, with faster-growing
solution given by
\begin{equation}
\sigma_{+}\left(k\right)=\frac{1}{2}\left(-\tau+\sqrt{\tau^{2}-4\Delta}\right).\label{eqn: quadratic-solution-of-sigma-PS}
\end{equation}
However, the addition of the term $-\varepsilon\nabla^{4}\phi$ to
Eq.(\ref{eqn: linear-concentration-evolution-PS}) leads to new definitions
for $\tau$ and $\Delta$:

\begin{eqnarray}
\tau\left(k\right) & = & \alpha+\left(C+B^{\prime}\right)k^{2}+\left(D+E^{\prime}\right)k^{4}\label{eqn: tau-PS}\\
\Delta\left(k\right) & = & \left(CA^{\prime}-AC^{\prime}\right)k^{2}+\left(CB^{\prime}+DA^{\prime}\right)k^{4}+\left(CE^{\prime}+DB^{\prime}\right)k^{6}+\left(DE^{\prime}\right)k^{8}.\label{eqn: Delta-PS}
\end{eqnarray}

An investigation of the limit $k\to\infty$ yields an altered expression
for the behavior of the very-smallest wavelengths 
\begin{equation}
\lim_{k\to\infty}\sigma_{+}\left(k\right)\approx-2E^{\prime}k^{4}-2\left(DB^{\prime}\right)k^{2}+\mathcal{O}\left(1\right).\label{eqn: large-k-limit-PS}
\end{equation}
This justifies the new term $-E^{\prime}\nabla^{4}\phi$ that we have
added to Eqn.(\ref{eqn: linear-concentration-evolution-BH}), for
without it, the smallest wavelengths would all be unstable in the
regime of phase separation ($B^{\prime}<0$). In contrast, the limit
$k\to0$ yields the same result as before:
\begin{equation}
\lim_{k\to0}\sigma_{+}\left(k\right)\approx-\frac{1}{A^{\prime}}\left(CA^{\prime}-AC^{\prime}\right)k^{2}.\label{eqn: small-k-limit-PS}
\end{equation}
This is important because the coefficients $\left\{ A,C,A^{\prime},C^{\prime}\right\} $
are associated only with the kinetic effects of the collision cascade,
and are unchanged by the generalization of the chemical diffusion
process. So, for instance, estimates of the values of these parameters
performed with the Bradley-Shipman theory in mind remain valid in
the present case.

Despite the new parameter $\varepsilon$ and the expanded range of
possible values for $B^{\prime}$, the possibilities for an instability
remain extremely limited. We recall that the system is unstable only
if one of $\tau$ or $\Delta$ is negative in Eqs.(\ref{eqn: tau-PS})-(\ref{eqn: Delta-PS}),
and that $A^{\prime}$, $D$, and $E^{\prime}$ are positive by definition.
Assuming further, as discussed above, that both $C$ alone, and also
the parameter group $\left(CA^{\prime}-AC^{\prime}\right)$, are positive,
the only terms in Eqs.(\ref{eqn: tau-PS})-(\ref{eqn: Delta-PS})
for this system that are not absolutely positive are those containing
$B^{\prime}$. We conclude that an instability can only occur for
a sufficiently negative value of $B^{\prime}$. Furthermore, because
the estimated positive value of $\left(CA^{\prime}-AC^{\prime}\right)$
stabilizes the long-wavelength modes near $k=0$, the bifurcation
would be of a finite-wavelength type, leading to the narrow band of
unstable modes typical of well-ordered systems.

\subsection{Full Bifurcation Criteria}

To obtain a more quantitative treatment of the instability criteria,
we first recall that Equation~(\ref{eqn: quadratic-solution-of-sigma-PS})
is stable is stable if both $\tau>0$ and $\sigma>0$, but unstable
if either one is negative. Hence, a bifurcation from stability to
instability occurs when either
\begin{equation}
\min_{k}\tau\left(k\right)=0\label{eqn: oscillatory-bifurcation}
\end{equation}
or 
\begin{equation}
\min_{k}\Delta\left(k\right)=0.\label{eqn: stationary-bifurcation}
\end{equation}
It is sufficient to examine these two criteria independently. As described
elsewhere, both in general \cite{cross-greenside-2009-book} and in
the context of ion irradiation \cite{abrasonis-morawetz-PRB-2012},
these two possibilities produce different kinds of instability: if
$\tau\left(k\right)>0$ but $\min_{k}\Delta\left(k\right)\le0$, the
growth rate is real and the instability is stationary; however, if
$\Delta\left(k\right)>0$ but $\min_{k}\tau\left(k\right)\le0$, the
growth rate is complex and the instability is oscillatory. As there
is no experimental evidence of oscillatory instabilities, we will
focus exclusively here on the stationary bifurcation criteria (\ref{eqn: stationary-bifurcation}).

To analyze the condition (\ref{eqn: stationary-bifurcation}), we
first, for simplicity, rename coefficients in Eq.\ref{eqn: Delta-PS}
to
\begin{equation}
\begin{aligned}a & =\left(CA^{\prime}-AC^{\prime}\right)\\
b & =\left(CB^{\prime}+DA^{\prime}\right)\\
c & =\left(CE^{\prime}+DB^{\prime}\right)\\
d & =\left(DE^{\prime}\right)
\end{aligned}
\label{eq: simple-coeff-defs}
\end{equation}
We also define
\begin{equation}
\tilde{\Delta}=\frac{\Delta}{k^{2}}=a+bk^{2}+ck^{4}+dk^{6}.\label{eq: delta-tilde}
\end{equation}
Now, the most unstable mode $k_{\text{max}}^{2}$ satisfies 
\[
\Delta^{\prime}\left(k\right)=\left(k^{2}\tilde{\Delta}\right)^{\prime}=k^{2}\tilde{\Delta}^{\prime}+2k\tilde{\Delta}=0,
\]
and a bifurcation occurs when 
\[
\Delta\left(k_{\text{max}}^{2}\right)=k_{\text{max}}^{2}\tilde{\Delta}\left(k_{\text{max}}^{2}\right)=0
\]
Because we are looking for $k_{\text{max}}^{2}>0$, we can therefore
assume that at the bifurcation to instability both 
\begin{eqnarray}
\tilde{\Delta}^{\prime} & = & 0\label{eqn: dtp-zero}\\
\tilde{\Delta} & = & 0\label{eqn: dt-zero}
\end{eqnarray}
From condition (\ref{eqn: dtp-zero}) we have
\begin{equation}
2k\left(b+2ck^{2}+3dk^{4}\right)=0\label{eq: quadratic-equation}
\end{equation}
to which we apply the quadratic formula to obtain the definition of
$k_{\text{max}}^{2}$:
\begin{equation}
k_{\text{max}}^{2}=\frac{-c+\sqrt{c^{2}-3bd}}{3d},\label{eqn: kmax-def}
\end{equation}
Here the positive root can be shown to always yield either the only
positive value for $k_{\mathrm{max}}^{2}$, or the value representing
a minimum of $\tilde{\Delta}$. Now, before inserting the result (\ref{eqn: kmax-def})
into Eq.(\ref{eqn: dt-zero}), we simplify the latter by means of
the result (\ref{eq: quadratic-equation}); this gives a simplified
version of Eq.(\ref{eqn: dt-zero}) in the form
\begin{equation}
3\tilde{\Delta}\left(k_{\text{max}}^{2}\right)=3a+2bk_{\text{max}}^{2}+ck_{\text{max}}^{4}=0.\label{eqn: pre-bifurcation}
\end{equation}
Finally, inserting Eqn.(\ref{eqn: kmax-def}) into the simplified
Eq.(\ref{eqn: pre-bifurcation}), eliminating the square root, and
simplifying the expression as much as possible, we obtain for the
stability boundary the final result
\[
3\left(ab-cd\right)^{2}+4\left(c^{2}-b^{2}\right)\left(ac-bd\right)=0.
\]
Unfortunately, this criteria is not particularly illuminating. It
is expressed in terms of the constants in Eq.(\ref{eq: simple-coeff-defs}),
which are themselves expressed in terms of environmental parameters
via Eqs.(\ref{eq: ACD})-(\ref{eq: alpha-beta-gamma}), which, \emph{in
turn}, depend upon the underlying tunable parameters of ion energy,
flux, etc.. In this situation, we therefore look for a relevant limiting
case of the general problem.

\subsection{The limit of weak coupling: $AC^{\prime}\ll CA^{\prime}$\label{sub: small-gamma-limit}}

In Ref.~\cite{norris-etal-arXiv-2013-coefficient-measurements},
we observed that for the highly-studied GaSb system, the parameter
group $AC^{\prime}$ was much smaller the parameter group $CA^{\prime}$
with the same dimensions. In fact, because the only place either $A$
or $C^{\prime}$ appear is in the first term $\left(CA^{\prime}-AC^{\prime}\right)$
of Eq.(\ref{eqn: Delta-PS}), the relative insignificance of the $AC^{\prime}$
term suggests considering, as a leading order approximation, the simplified
equations 
\begin{eqnarray}
\frac{\partial h}{\partial t} & = & C\nabla^{2}h-D\nabla^{4}h\label{eqn: small-gamma-reduction-h}\\
\frac{\partial\phi}{\partial t} & = & -A^{\prime}\phi+B^{\prime}\nabla^{2}\phi-E^{\prime}\nabla^{4}\phi,\label{eqn: small-gamma-reduction-phi}
\end{eqnarray}
in which both the term $-A\phi$ of Eq.(\ref{eqn: linear-height-evolution-PS}),
and the term $C^{\prime}\nabla^{2}h$ of Eq.(\ref{eqn: linear-concentration-evolution-PS}),
have been neglected. Critically, we observe that in this approximation,
the equations have completely decoupled. Of course, a small correction
to the solution of these equations preserves the coupling and so prevents
the two fields from evolving completely independently. However, this
correction would be expected to have relative magnitude $\frac{AC^{\prime}}{CA^{\prime}}\ll1$,
and so we label this limit as the \emph{limit of weak coupling}. 

The linear stability of this simplified system is quite simple; inserting
the standard ansatz (\ref{eqn: infinitesimal-perturbation}) into
Eqs.(\ref{eqn: small-gamma-reduction-h})-(\ref{eqn: small-gamma-reduction-phi}),
we find the two solutions for the dispersion relation satisfy
\begin{eqnarray}
\sigma_{1}\left(k\right) & = & -Ck^{2}-Dk^{4}\label{eq: simple-dispersion-height}\\
\sigma_{2}\left(k\right) & = & -A^{\prime}-B^{\prime}k^{2}-E^{\prime}k^{4}.\label{eq: simple-dispersion-concentration}
\end{eqnarray}
However, because both $C$ and $D$ are assumed positive in our regime,
the first growth rate is negative for all wavenumbers, and the stability
of the system is entirely determined by the second solution (\ref{eq: simple-dispersion-concentration})
associated with concentration. There, if $B^{\prime}$ is negative,
the most unstable mode $k_{\text{max}}$ of this approximate model,
defined via $\sigma^{\prime}\left(k_{\text{max}}\right)=0$, is

\begin{equation}
k_{\text{max}}=\left(\frac{-B^{\prime}}{2E^{\prime}}\right)^{\frac{1}{2}}=\left(\frac{-F^{\prime\prime}\left(c_{0}\right)}{2\tilde{\varepsilon}}\right)^{\frac{1}{2}},\label{eq: critical-wavenumber}
\end{equation}
which gives a most unstable wavelength of 
\[
\lambda_{\text{max}}=2\pi\sqrt{\frac{2\tilde{\varepsilon}}{-F^{\prime\prime}\left(c_{0}\right)}}.
\]
If one takes a very crude model of the free energy of mixing
\[
F_{\mathrm{mix}}=H_{\mathrm{mix}}\left(c\right)-TS_{\mathrm{mix}}\left(c\right),
\]
with the enthalpy of mixing $H_{\mathrm{mix}}$ a strongly convex
function of concentration near the steady state $c_{0}$ (as in Figure~(\ref{fig: phase-diagram-and-free-energy})b),
and the entropy of mixing $S_{\mathrm{mix}}$ obeying a standard solution
model that is more weakly convex,
\[
H_{\mathrm{mix}}^{\prime\prime}\left(c_{0}\right)<S_{\mathrm{mix}}^{\prime\prime}\left(c_{0}\right)<0,
\]
 then the denominator $-F^{\prime\prime}\left(c_{0}\right)$ inside
the square root is a decreasing function of the temperature $T$.
Provided the temperature (or its effective equivalent \cite{martin-PRB-1984})
grows with ion energy, then the model predicts an increasing wavelength
as the ion energy increases, which is at least qualitatively consistent
with existing experimental results \cite{facsko-etal-PRB-2001}.

Turning to the bifurcation criteria, defined via $\sigma\left(k_{\text{max}}\right)>0$,
we have simply
\begin{equation}
\frac{\left(B^{\prime}\right)^{2}}{4E^{\prime}}>A^{\prime},\label{eq: small-gamma-bifurcation-criteria}
\end{equation}
or in physical parameters,
\begin{equation}
\frac{\tilde{D}}{4\tilde{\varepsilon}}\frac{\Omega}{\Delta}\left[F^{\prime\prime}\left(c_{0}\right)\right]^{2}>P_{0}\left[c_{B,b}\Lambda_{A}^{\prime}\left(c_{A,0}\right)+c_{A,b}\Lambda_{B}^{\prime}\left(c_{B,0}\right)\right].\label{eq: instability-criteria-binary}
\end{equation}
Here, on the left-hand side of the inequality, $F\left(c_{0}\right)$
and $\tilde{\varepsilon}$ are associated with the energetics (driving
force) of the binary material, and $\tilde{D}$ with the kinetics
(mobility). On the right-hand side, the expanded parameter $A^{\prime}$
describes the replenishment of species $A$ from the bulk, minus the
sputter rate of species $A$ at the interface; it thus represents
an ion-energy dependent \emph{net replenishment rate}, which is proportional
to the net sputter rate. Unfortunately, many of the parameters in
the inequality (\ref{eq: instability-criteria-binary}) vary non-trivially
with ion energy, and more study is needed to extract precise predictions.
However, qualitatively, the instability criteria is clearly a competition
between chemical energetics and kinetics on the one hand, and ion
erosion on the other. Even more simply, the criteria can be expressed
as a comparison of two \emph{timescales}: instability will occur only
if the amorphous layer has enough time to phase separate before it
is sputtered away and replaced with fresh material from the bulk.

\subsection{Unique Role of the ion Beam. }

It is worth taking special note of three unique contributions of the
ion beam to the regime described here, relative to more typical thermally-driven
phase separation systems. First, the makes phase separation energetically
possible: by preferentially sputtering one component of a common binary
material, it drives the system from an energetically-stable stoichiometry
to an energetically-unstable one. Second, it makes phase separation
possible kinetically. Normally, even in an unstable stoichiometry,
the film would not phase separate at room temperature. However, the
ion beam supplies a great deal of noise into the film, giving atoms
the mobility to travel down gradients of the chemical potential. 
\begin{figure}
\centering{}\includegraphics[width=3in]{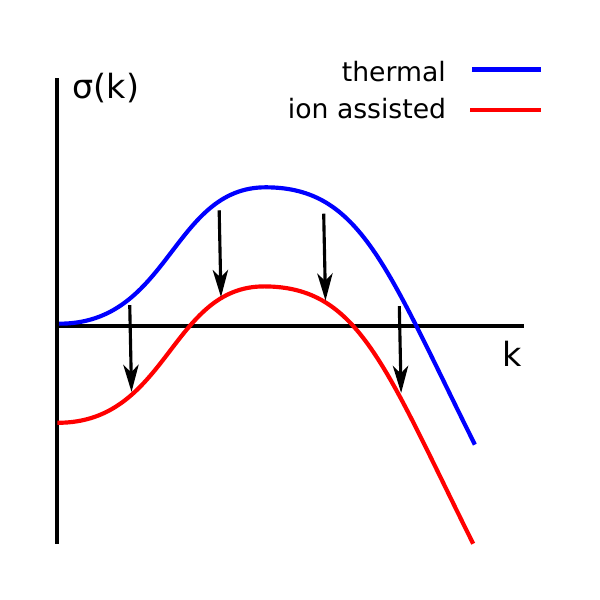}\caption{An schematic illustration of the connection between ion assisted phase
separation and thermal phase separation, revealed in the limit of
weak coupling $A\gamma\ll C\alpha$. Here, the effect of material
replenishment is subtract a constant from the dispersion relation,
which shifts the entire curve downward. This turns the longwave instability
typical of thermal phase separation into one with a narrow band of
unstable modes.\label{fig: longwave-to-shortwave}}
\end{figure}

Finally, and most striking, the ion beam transforms phase separation
from a longwave into a finite-wavelength process, as illustrated in
Fig.\ref{fig: longwave-to-shortwave}. In a standard, thermally-driven
phase separation system, one has the Cahn-Hilliard equation \cite{cahn-hilliard-1958:ch-equation},
for which the linearization yields
\begin{equation}
\frac{\partial\phi}{\partial t}=-B^{\prime}\nabla^{2}\phi-E^{\prime}\nabla^{4}\phi.\label{eqn: Cahn-Hilliard}
\end{equation}
This equation is unstable, but the instability is \emph{longwave}:
i.e., the long-wavelength modes near $k=0$ are unconditionally unstable.
Because of this feature of the instability, thermal phase separation
generically exhibits disordered structures and coarsening, under the
well-known Ostwald Ripening phenomenon \cite{ostwald-1900:ripening}.
In contrast, under ion-irradiation, fresh material, at the bulk concentration,
is always entering the film as the old material is sputtered away
\cite{bradley-shipman-PRL-2010,bradley-PRB-2011c}. This process contributes
the leading-order damping term $\phi_{t}\propto-A^{\prime}\phi$ in
Eqn.(\ref{eqn: small-gamma-reduction-phi}), which stabilizes the
long waves and prevents coarsening, enabling the emergence of ordered
patterns. Ironically, this term is similar to one conjectured to exist
a decade ago by the original observers of ordered dots in GaSb \cite{facsko-etal-PRB-2004},
except that it is located in the equation for the concentration field
rather than that of the height field. Indeed, it was obtained even
earlier to describe phase separation during deposition \cite{atzmon-kessler-srolovitz-JAP-1992},
in an envisioned limit of surface confined atomic mobility. Ion beams
may thus be seen as a tool able to achieve this intriguing theoretical
limit.

\section{A Distinguishing Experimental Signature\label{sec: experimental-signature}}

Any proposed alternative to an existing theory must be testably distinguishable
from the original. We therefore provide here an experimental signature
that distinguishes between the morphological instability of Bradley
and Shipman \cite{bradley-shipman-PRL-2010,shipman-bradley-PRB-2011},
and the chemically-driven instability just proposed. The signature
arises from the structure of the matrix in Eq.(\ref{eqn: stability-matrix-BS})
describing the linear evolution of perturbations, which we will examine
through the lens of each theory in turn. It will be particularly important
in what follows to carefully recall the definition of the coefficient
$A$ in Eq.(\ref{eq: ACD}). This coefficient is often described as
representing ``preferential sputtering.'' However, the phrase ``preferentially
sputtered'' is most naturally interpreted as a statement on the relative
values of $\Lambda\left(c\right)$: i.e., the material with the greater
value of $\Lambda$ is preferentially sputtered, and leads to enrichment
of the opposing component. In contrast, the constant $A$ contains
the \emph{derivatives} $\Lambda^{\prime}\left(c\right)$ of the sputter
yields. It therefore characterizes the effect of changing concentration
on the total erosion rate. It is entirely possible for species $A$
to have a larger yield than species $B$, and hence to be ``preferentially
sputtered,'' but simultaneously to induce \emph{less} total erosion
when its concentration is increased. 

\begin{figure}
\centering{}\includegraphics[width=3in]{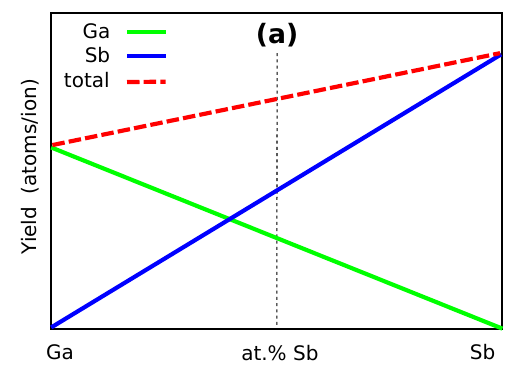}\includegraphics[width=3in]{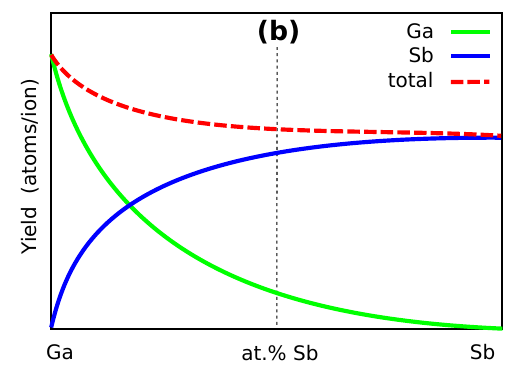}\caption{Comparison of two possible idealized yield curves $\Lambda_{A}\left(c\right)$,
$\Lambda_{B}\left(c\right)$. In both figures, Sb is ``preferentially
sputtered'' relative to Ga at the 50 at\% concentration. However,
the net sputtering has a qualitatively different dependence on concentration.
Schematic (a) would be associated with an idealized compound for which
the yields are linear in the concentration. Here, increasing antimony
concentrations always lead to increased sputter rates. However, schematic
(b) would describe a system subject to surface enrichment, which appears
as nonlinear dependence of the yield on concentration. Here, despite
the fact that antimony is preferentially sputtered at 50 at.\%, increasing
the antimony concentration \emph{decreases }the overall yield.\label{fig: yield-scenario-comparison}}
\end{figure}

This possibility is illustrated in Figure~\ref{fig: yield-scenario-comparison}
by two potential, idealized yield curves for the system Ar$\to$GaSb
(the true curves are unknown), where one observes Sb to be preferentially
sputtered during the early stages of irradiation. In the first sketch,
we illustrate the possibility that the concentration dependence of
the yields of Ga and Sb are assumed to be linearly proportional to
concentration. Hence, because Sb is preferentially sputtered from
a target at the 50/50 composition, then increasing the total amount
of Sb increases the total sputter rate, and we would choose Sb to
be species A. An alternative possibility is envisioned in the second
sketch, inspired by the observation that GaSb seems to undergo Gibbsian
surface segregation \cite{yu-sullivan-saied-SS-1996-gibbsian-segregation,el-atwani-etal-JAP-2011},
in which a very thin surface layer becomes rich in Sb due to its lower
surface energy relative to Ga. One can easily envision that this would
enhance the sputtering of Sb across a wide range of concentrations,
which manifests itself as a nonlinearity in the curves. Here, although
Sb is ``preferentially sputtered'' across a wide range of concentrations,
increasing the Sb concentration actually \emph{decreases} the overall
sputter rate, and so we would choose Ga to be species A.

\begin{figure}
\begin{centering}
\includegraphics[width=5in]{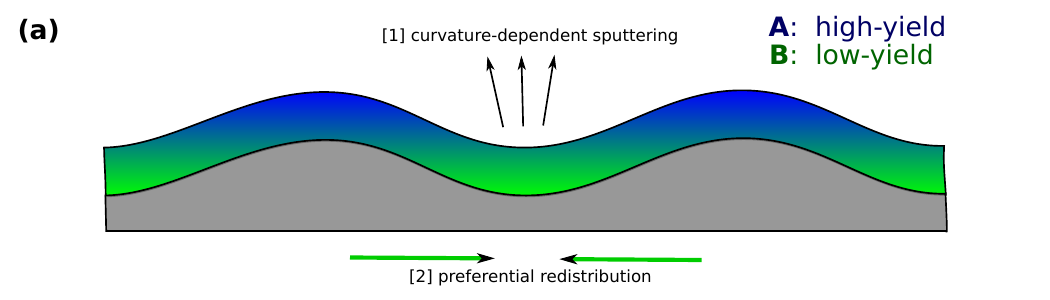}\vspace{1cm}

\par\end{centering}

\centering{}\includegraphics[width=5in]{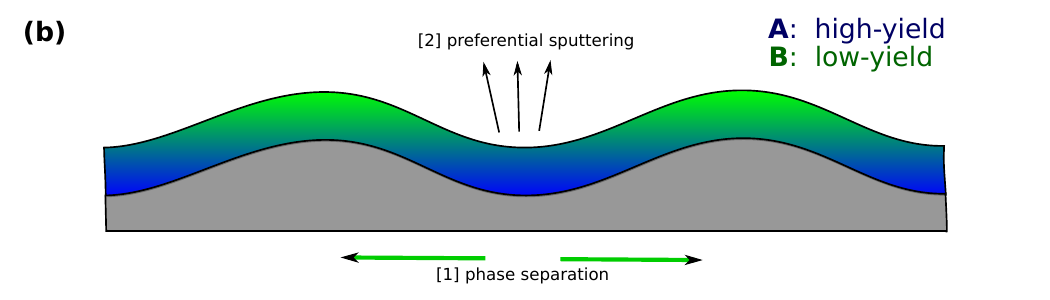}\caption{Representative illustrations of two finite-wavelength instability
regimes. (a) In a kinetically-driven instability, the surface geometry
is fundamentally unstable, while the chemistry follows the interface
via the preferential redistribution term $C^{\prime}\nabla^{2}u$
in Eqn.(\ref{eqn: linear-concentration-evolution-PS}). To stabilize
long waves, it is necessary that the low-yield species B be preferentially
pushed into the valleys, which leaves the hilltops saturated with
the high-yield species A. (b) In a chemically-driven instability,
the surface is fundamentally stable, following the unstable chemistry
via the preferential sputtering term $-A\phi$ in Eqn.(\ref{eqn: linear-height-evolution-PS}).
Hence, the hilltops are left saturated with the low-yield species
B. Mass redistribution acts to stabilize against long waves. \label{fig: instability-schematics}}
\end{figure}

With this clarification, we now consider the theory of Bradley and
Shipman, to which the second row of the matrix (\ref{eqn: stability-matrix-BS})
is relevant: 
\begin{equation}
C^{\prime}k^{2}u_{1}+\left(\sigma\left(k\right)+A^{\prime}+B^{\prime}k^{2}\right)\phi_{1}=0\label{eq: matrix-line-two-BS}
\end{equation}
Here $\sigma>0$ characterizes any instability, $A^{\prime}>0$ by
definition (replenishment), and $B^{\prime}>0$ by assumption (simple
Fickian diffusion). Hence, the relative phases of the height and concentration
fields (as determined by the relative signs of $u_{1}$ and $\phi_{1}$)
are determined by the sign of the coefficient $C^{\prime}$. Because
the Bradley-Shipman instability requires that $C^{\prime}<0$, we
conclude that $u_{1}\phi_{1}>0$, so that the concentration variations
of species $A$ are in phase with the height variations. The physical
interpretation of this scenario is that the surface first undergoes
undulations due to an erosive instability, after which the preferentially-redistributed
species is driven into the valleys \cite{bradley-shipman-PRL-2010}.
The negative value of the product $AC^{\prime}$ implies that the
species producing a higher net sputter yield is left on the hilltops,
which stabilizes the longwaves (see Figure \ref{fig: instability-schematics}a).

We next turn to the chemically-driven instability proposed here, to
which the first row of the matrix (\ref{eqn: stability-matrix-BS})
is relevant (this row is not changed by the generalization we have
performed):
\begin{equation}
\left(\sigma\left(k\right)+Ck^{2}+Dk^{4}\right)u_{1}+A\mbox{\ensuremath{\phi}}_{1}=0.\label{eq: matrix-line-one-SN}
\end{equation}
Here, again, $\sigma>0$ characterizes any instability, we are assuming
$C>0$ as discussed above, $D>0$ by definition, and $A>0$ by choice
of species A. Hence, we conclude that $u_{1}\phi_{1}<0$, so that
the concentration variations are now \emph{out of phase} with the
height variations. A physical interpretation of this scenario, as
suggested in the numerical study of Le Roy et al. \cite{le-roy-etal-PRB-2010-phase-separation},
is that preferential sputtering first induces the material to phase-separate
into large regions of 50/50 GaSb containing small islands of the species
producing a lower net sputter yield, which then recede more slowly
than the bulk of the surface (see Figure \ref{fig: instability-schematics}b).

Because of this notable experimental difference between the two mechanisms,
it would be invaluable to determine experimentally both the relative
yields of Ga and Sb in terms of the composition of the amorphous film,
so as to identify the species leading to greater overall sputtering,
and then to identify the relative phase of that species relative to
the height field during the early stages of pattern formation. Unfortunately,
although it is known that Sb sputters preferentially from a fresh
50/50 target, the relative slopes of the yield curves with respect
to concentration are currently unknown, as are the relative concentrations
at the peaks and troughs of the structures. Using ex-situ electron
energy loss spectroscopy (EELS) applied to a \emph{single} nanostructure,
Le Roy et al. observe very well-defined Ga-rich caps at the top of
the structure \cite{le-roy-etal-JAP-2009}. However, those observations
were performed in the presence of oxygen, which preferentially reacts
with Ga, and can therefore produce results that differ significantly
from in-situ conditions \cite{el-atwani-etal-JAP-2011}. Therefore,
both sets of data needed to distinguish between the kinetic and chemical
routes to instability remain (important) open experimental questions.

\paragraph*{}

\section{Conclusions}

Motivated by a growing body of experimental \cite{fukutani-etal-JJAP-2008,le-roy-etal-JAP-2009,abrasonis-etal-JAP-2010,hoffsass-etal-APA-2012}
and numerical \cite{fukutani-etal-JJAP-2008,le-roy-etal-PRB-2010-phase-separation}
evidence suggesting that phase separation plays an important role
in pattern formation during the irradiation of two-component targets,
and molecular dynamics simulations suggesting that existing theories
neglecting this phenomenon may not explain the patterns \cite{norris-etal-arXiv-2013-coefficient-measurements},
we have presented a generalization of the Bradley-Shipman theory of
irradiated binary compounds \cite{bradley-shipman-PRL-2010,shipman-bradley-PRB-2011}
that admits the phase separation mechanism. We demonstrated that the
resulting model readily admits the development of ordered patterns
even for parameter values where the BS theory cannot. A special limit
suggested by the estimated parameter values allows a simple, intuitive
understanding of the instability criteria, and highlights the potentially
unique nature of ion-assisted phase separation relative to thermal
phase separation. Finally, we identified a distinct experimental signature
that distinguishes between the two instabilities, that highlights
the need for specifica additional experimental data on this system.
\begin{acknowledgments}
The authors thank R. Mark Bradley, Gintas Abrasonis, Mathis Plapp,
and Jean-Paul Allain for helpful discussions. 
\end{acknowledgments}
\bibliographystyle{unsrt}
\bibliography{/home/snorris/Dropbox/research/bibliography/tagged-bibliography,/home/scott/Dropbox/research/bibliography/tagged-bibliography}

\end{document}